\documentstyle[11pt,paspconf,epsfig]{article}

\begin{document}

\title{
What can we learn from the fluxes of the 1.2 Jy IRAS sample?    }

\author{St\'ephane Rauzy and Martin Hendry}
\affil{Dept. of Physics and Astronomy, University of Glasgow, UK}





\begin{abstract}
We present a new method for fitting peculiar velocity models
to complete flux limited magnitude-redshifts catalogues, using
the luminosity function of the sources as a distance indicator.
The method is characterized by its robustness.
In particular, no assumptions are made concerning
the spatial distribution of sources and their luminosity
function. Moreover the inclusion of additional observables,
such for example the one carrying the Tully-Fisher information,
is straightforward.
\\
As an illustration of the method,
the predicted IRAS peculiar velocity model
is herein tested
using the fluxes of the IRAS 1.2 Jy sample as the distance indicator.
The results suggest that this model,
while successful
in reproducing locally the cosmic flow, fails to describe the
kinematics on larger scales.
\end{abstract}




%
%

\section{The method}

The application of the method is restricted to
samples strictly complete up to
a given magnitude limit $m_{\rm lim}$, 
i.e. the selection function in apparent magnitude is well described 
by a sharp cut-off
$\psi(m)=\theta(m_{\rm lim}-m)$ with
$\theta(x)$ the Heaveside function. 
The probability density of the sample may be written in this case as
\begin{equation}\label{dP_2}
dP = \frac{1}{A} h(\mu,l,b) \cos b \,dl db d\mu \, f(M) dM
\, \theta(m_{\rm lim}-m)
\end{equation}
where $\mu=m-M$ is the distance modulus, $h(\mu,l,b)$ the line-of-sight
distribution function, $f(M)$ the luminosity function  
and $A$ is the normalisation
factor warranting $\int dP =1$. 

The milestone of the method consists in defining the random
variable $\zeta$ as follows
\begin{equation}\label{zeta}
\zeta = \frac{F(M)}{F(M_{\rm lim})}
\,\,\,\,\,\,\,
\,\,\,\,\,\,\,;
\,\,\,\,\,\,\,
\,\,\,\,\,\,\,
d\zeta = \frac{f(M)}{F(M_{\rm lim})} \,dM
\end{equation}
where $F(M)=\int_{-\infty}^M f(x)dx$ stands for the cumulative
distribution function in $M$ and $M_{\rm lim}\equiv
M_{\rm lim}(\mu)$ is the maximum absolute magnitude for which
a galaxy at distance $\mu$ would be visible in the sample
(e.g.  $M_{\rm lim}(\mu)=m_{\rm lim} - \mu$ if the k-correction is
neglected). By definition, the random variable $\zeta$ for a
sampled galaxy belongs to the interval $[0,1]$. The probability
density of Eq. (\ref{dP_2}) reduces to
\begin{equation}\label{dP_3}
dP =  \frac{1}{A} h(\mu,l,b)\,F(M_{\rm lim}(\mu))
\,\cos b\,dldb d\mu \, \times \,
\theta(\zeta)
\, \theta(1-\zeta)
\, d\zeta
\end{equation}
This equation implies
the two following properties:
\begin{itemize}
\item P1: $\zeta$ is uniformly distributed between $0$ and $1$.
\item P2: $\zeta$ and $\mu$ are statistically independent,
i.e. the distribution of $\zeta$ does not depend on
the spatial position of the galaxies.
\end{itemize}


The random variable $\zeta$ can be estimated without any prior
knowledge of the cumulative luminosity function $F(M)$. For
each galaxy $(M_i,\mu_i)$ one can indeed form the subsample
$S_i=S_1 \cup S_2$ with
$S_1 =  \{\,(M,\mu) {\rm ~such~that~}
 M \le M_i {\rm ~and~} \mu \le \mu_i \,\}$
and $S_2 =  \{\,(M,\mu) {\rm ~such~that~} M_i <  M \le M^i_{\rm lim}
 {\rm ~and~} \mu \le \mu_i \,\}$.
By construction (see figure 1)
$M$ and $\mu$ are independent in each subsample $S_i$. This implies
that the following quantity is an unbiased estimate of the
random variable $\zeta$
\begin{equation}\label{zetaestimate}
{\hat \zeta_i} =  \frac{r_i}{n_i+1}
\end{equation}
where $n_i$ is the number of objects in
$S_i=S_1 \cup S_2$ and $r_i$ the number of objects in $S_1$
(see Efron \& Petrosian 1992).



The radial peculiar velocity field is herein described by a
linear model parametrized by a $N$-dimensional
vector ${\bf \beta}=(\beta_1,\beta_2,...,\beta_N)$,
\begin{equation}\label{velocitymodel}
v_{\bf \beta} \equiv v_{\bf \beta}({\bf r}) =
\sum_{k=1}^N \beta_k \,v_k({\bf r})
\end{equation}
where $v_1({\bf r})$, $v_2({\bf r})$, ..., $v_N({\bf r})$ is a
set of functions depending on the spatial position ${\bf r}$.
It assumed hereafter that there exists a solution ${\bf \beta}^{\star}$
fairly reproducing the true radial peculiar velocities of
galaxies $v_i \equiv v({\bf r}_i)\simeq  v_{{\bf \beta}^{\star}}({\bf r}_i)$.
For a given vector ${\bf \beta}$, the model dependent variables
$\mu_{\bf \beta}$ and
$M_{\bf \beta}$ can be computed from the observed redshift $z$ and
apparent magnitude $m$ following
\begin{equation}\label{mubeta}
\mu_{\bf \beta}= 5\,\log_{10} \frac{cz}{H_0} + 25 -
u_{\bf \beta}
\,\,\,\,;
\,\,\,\,
M_{\bf \beta}= m - \mu_{\bf \beta}
\,\,\,\,;
\,\,\,\,
u_{\bf \beta}= - 5\,\log_{10} 
\left (1 - \frac{v_{\bf \beta} }{cz} \right )
\end{equation}
The quantities
$\mu_{\bf \beta}$ and
$M_{\bf \beta}$ are related to the true absolute magnitude $M$ and distance
modulus $\mu$ via
$
\mu_{\bf \beta} = \mu  +
u_{\bf \beta^{\star}} -
u_{\bf \beta}$ and
$M_{\bf \beta}
= M
- u_{\bf \beta^{\star}} +
u_{\bf \beta}$.

It is shown in Rauzy\&Hendry (hereafter RH)
 that any test of independence between
$\zeta_\beta$ computed following
Eq. (\ref{zetaestimate})
and
$u_\beta$
provides us with an unbiased estimate of $\beta^{\star}$.
In particular the coefficient of correlation
$\rho(\zeta_\beta,u_\beta)$ has to vanish when
$\beta  = {\beta^{\star}}$, i.e.
\begin{equation}\label{Correlationcoefficient}
\beta  = {\beta^{\star}}
\,\,\,\,\,
\Longleftrightarrow
\,\,\,\,\,
\rho(\zeta_\beta,u_\beta)=0
\end{equation}
The accuracy of this estimator can be obtained through
numerical simulations by analysing the influence
of sampling fluctuations on the coefficient of correlation.

This estimator is clearly robust. 
Nothing has been assumed concerning the spatial distribution of
sources nor on the shape of their luminosity function. In
particular the method is free of Malmquist-like biases.
Moreover 
the inclusion of
a second observable parameter, for example in order to account
for Tully-Fisher information, can be achieved without any
difficulty. Finally it is worthwhile to mention that
the method will
benefit from the use of the velocity orthogonalization procedure
proposed by Nusser \& Davis (1995) (see RH for details).
\begin{figure}
\begin{center}
{\epsfig{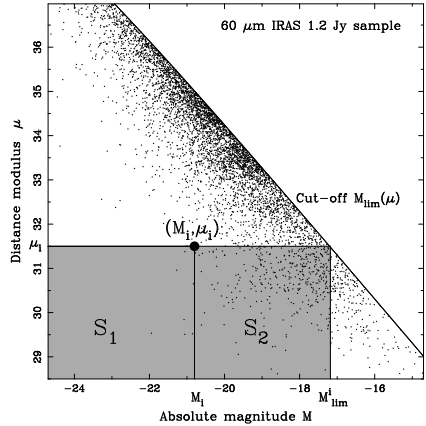}}
\caption{
Distance modulus versus absolute magnitude for the 60 $\mu$m
IRAS 1.2 Jy sample (5321 galaxies)
}
\end{center}
\end{figure}
\section{Application to the IRAS 1.2 Jy sample}

The method is herein applied to the 60 $\mu$m IRAS 1.2 Jy sample
(Fisher et al. 1995). 
The distance modulus versus absolute magnitude diagram is
shown in figure 1.
The peculiar velocity field model tested is the
predicted IRAS velocity field
(Strauss et al. 1992)
characterized by only one parameter $\beta=\Omega_{0.6}/b_I$.
%
The luminosity function
of these sources does not exhibit any turnover towards the
faint-end tail, at least within the observed range of magnitudes.
Due to the large spread of the LF, one
cannot expect very high constraints on the velocity model tested.
However a 
%
 rejection test for the $\beta$ parameter
can be constructed. We obtained that 
$\beta \ge 0.7$ can be rejected with a confidence level of
95\% and $\beta \ge 1.1$ with a confidence level of
99\%.

In a second step, we use the observed correlation between
the absolute magnitude $M$ and some "colour index" defined
as $p=2.5 \log_{10} (F_{100}/F_{60})$ 
(with $F_{100}$ the
flux at 100 $\mu$m) in order to refine the analysis. The data
have been grouped in $8$ classes by interval of $p$. 
Because of the correlation between $p$ and
$M$, the spread of the luminosity function
for each of these classes taken individually is expected to be
smaller than the spread of the global luminosity function,
and thus the accuracy of the distance indicator improved.
The random variable $\zeta_\beta$ is then computed using
Eq. (\ref{zetaestimate}) but this time class by class.
The correlation
between $\zeta_\beta$ and
the velocity modulus $u_\beta$ is after that evaluated for the whole sample.

\begin{figure}
\begin{center}
{\epsfig{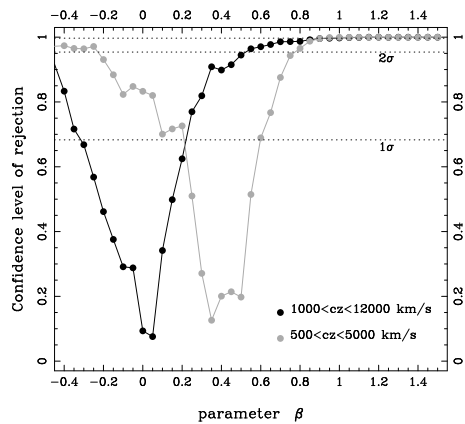}}
\caption{
Confidence level of rejection for the parameter $\beta$ 
}
\end{center}
\end{figure}
The results are presented in figure 2 in terms of the confidence level
of rejection for the parameter $\beta$.
The method has been first applied to the galaxies within the redshift range
1000-12000 km s$^{-1}$.  It is found that $\beta \in [-0.35,0.25]$ at
$1\sigma$, and that models with
$\beta \ge 0.5$ can be rejected with a confidence level of
95\%. This result is in disagreement with most of the
analyses based on Tully-Fisher data (e.g. VELMOD on MarkIII
(Willick \& Strauss 1998), ITF method on SFI (Da Costa et al. 1998)
favoring a value of $\beta = 0.5$.
We interpret this discrepancy as follows.

When fitting a velocity model to data, the natural weight assigned
by the fitting procedure
to each galaxy is roughly proportional to the inverse of its redshift.
The mean effective depth of the volume where the velocity model is
compared to data has to be estimated using these weights. For
our first sample with $z \in [1000,12000]$ km s$^{-1}$,  we find
a mean effective depth
$3800$ km s$^{-1}$.
In order to mimic the effective volume sampled by Tully-Fisher data
we have applied the method to a truncated version of the IRAS sample
containing 1621 galaxies with $z \in [500,5000]$ and galactic latitude
$|b| > 20$ (the mean effective depth of this sample is 2200 km s$^{-1}$).
Figure 2 shows that the value of $\beta$ estimated from this
truncated sample is fully consistent with the values obtained using
Tully-Fisher data. An interpretation of these results could
be that the predicted IRAS velocity field model, while successful
in reproducing locally the cosmic flow, fails to describe the
kinematics on larger scales.

{\acknowledgements {We are thankful to Michael Strauss for
providing us with the predicted IRAS peculiar velocity model.}}

\end{document}